\documentclass[conference]{IEEEtran}

\IEEEoverridecommandlockouts

\usepackage{cite}
\usepackage{amsmath,amssymb,amsfonts}
\usepackage{algorithm}
\usepackage{algorithmic}
\usepackage{graphicx}
\usepackage{textcomp}
\usepackage{xcolor}
\usepackage{url}
\usepackage{listings}
\usepackage{booktabs}
\usepackage{multirow}
\usepackage{array}
\usepackage{eso-pic}

\graphicspath{{figures/}{./}}

\newcommand{\tool}{\textsc{PatternDSE}}
\newcommand{\allo}{Allo}
\lstdefinestyle{allopy}{
  language=Python,
  basicstyle=\ttfamily\scriptsize,
  numbers=left,
  numberstyle=\tiny,
  stepnumber=1,
  numbersep=4pt,
  columns=fullflexible,
  keepspaces=true,
  frame=single,
  xleftmargin=1.2em,
  framexleftmargin=1.0em,
  breaklines=true,
  showstringspaces=false
}

\makeatletter
\def\ps@IEEEtitlepagestyle{%
  \def\@oddfoot{\mycopyrightnotice}%
  \def\@evenfoot{}%
}
\def\mycopyrightnotice{%
  {\footnotesize XXX-X-XXXX-XXXX-X/XX/\$XX.00~\copyright~20XX IEEE\hfill}%
  \gdef\mycopyrightnotice{}%
}
\makeatother

\newcommand\AtPageUpperMyright[1]{\AtPageUpperLeft{%
 \put(\LenToUnit{0.17\paperwidth},\LenToUnit{-2cm}){%
     \parbox{0.9\textwidth}{\raggedleft\fontsize{8}{11}\selectfont #1}}%
 }}%
\newcommand{\conf}[1]{%
\AddToShipoutPictureBG*{%
\AtPageUpperMyright{#1}
}
}

\begin{document}

\title{\vspace*{1cm}Pattern-Guided Design Space Exploration for FPGA Accelerator Design}
% If applicable, add funding information with:

\author{
\IEEEauthorblockN{Jialiang Zhang$^1$, Weiman Yan$^1$, Yuelin Zou$^2$}
\IEEEauthorblockA{$^1$Dept. of Electrical and Computer Engineering, University of Illinois Urbana-Champaign;\\ $^2$Department of Industrial Engineering and Operations Research, Columbia University; \\
$^1$Urbana, Illinois, USA\\ $^2$Essex Junction, Vermont, USA\\
email: jz23@illinois.edu}   %<------ Line breaks in the current column
}

\maketitle
\conf{\textit{Proc. of the International Conference on Electrical, Computer, Communications and Mechatronics Engineering (ICECCME 2026) \\
15--17 October 2026, Bali, Indonesia}}

\begin{abstract}
High-level synthesis (HLS) raises the abstraction level of FPGA accelerator design from hardware description languages to C/C++, but high-quality results still depend on schedule decisions such as pipelining, unrolling, tiling, reordering, and buffering. These decisions create a combinatorial design space, while many numerical kernels exhibit recurring computation patterns that suggest different optimization strategies.

This paper presents \tool{}, a lightweight pattern-guided design space exploration (DSE) framework for FPGA kernels written in \allo{}, a scheduling-oriented HLS programming system. \tool{} maps recurring computation patterns, including elementwise maps, reductions, matrix-vector operations, matrix-matrix operations, and stencil-like updates, to compact schedule spaces. It then applies candidate schedules, validates functional correctness through LLVM execution, checks HLS C code generation, and uses a simple pattern-aware estimator to rank candidates before Vitis HLS synthesis.

We evaluate \tool{} on six representative kernels: \texttt{vecadd}, \texttt{axpy}, \texttt{dot}, \texttt{matvec}, \texttt{gemm}, and \texttt{jacobi2d}. Compared with an exhaustive-lite baseline, pattern-guided DSE reduces the number of HLS-evaluated candidates from 140 to 29, achieving a 4.83$\times$ overall search reduction and up to 12.0$\times$ reduction for individual kernels. Across all evaluated kernels, \tool{} recovers the same best valid Vitis HLS latency as the exhaustive-lite baseline, demonstrating that computation-pattern information can prune unproductive schedule combinations while preserving high-quality HLS outcomes.
\end{abstract}

\begin{IEEEkeywords}
FPGA, High-level synthesis, Design space exploration, Optimization, Hardware accelerator, Software-Hardware Codesign
\end{IEEEkeywords}

\section{Introduction}

The growing diversity of modern workloads has made data-intensive applications, including machine learning, video processing, graph analytics, genomics, and layout generation, increasingly important across cloud and edge platforms~\cite{yan2025stiff,9434098,11132857,11240639}. Meanwhile, the slowdown of conventional processor scaling has increased the demand for domain-specific acceleration under tight power and area constraints~\cite{dark_silicon}. Field-programmable gate arrays (FPGAs) address this need by combining post-fabrication flexibility with customized datapaths, fine-grained parallelism, and application-specific memory organization. Consequently, FPGAs have been adopted in cloud-scale systems and datacenter services~\cite{putnam_catapult,caulfield_cloud_scale}, as well as a broad range of data-intensive application domains~\cite{hls_survey}.

High-level synthesis (HLS) reduces the programming burden by compiling C/C++ or higher-level descriptions into hardware implementations. However, HLS optimization remains highly dependent on architectural and scheduling choices, including pipelining, unrolling, tiling, reordering, memory partitioning, and buffering. Since these choices interact combinatorially, HLS optimization is fundamentally a design space exploration (DSE) problem rather than a push-button compilation task~\cite{hls_survey,schafer_hls_dse}.

Recent HLS programming systems expose schedules as first-class objects, allowing algorithms and hardware mappings to be developed separately. \allo{} follows this direction by providing a scheduling-oriented interface for generating multiple hardware implementations from a common kernel specification~\cite{allo}. While this improves programmability, it does not determine which schedules should be explored.

This paper argues that schedule exploration can be reduced and made more interpretable by exploiting computation structure. Numerical kernels are not arbitrary programs: elementwise maps, reductions, matrix-vector products, matrix-matrix products, and stencil-like updates exhibit different reuse patterns, dependency constraints, and memory bottlenecks. This pattern-oriented view is consistent with the Berkeley ``dwarfs'' and accelerator benchmark suites such as MachSuite, Rodinia, CHStone, Rosetta, and MLPerf, which organize workloads around recurring computational motifs~\cite{asanovic_dwarfs,machsuite,rodinia,chstone,rosetta,mlperf_inference}.

We present \tool{}, a lightweight pattern-guided DSE prototype for FPGA kernels written in \allo{}. \tool{} maps supported kernel patterns to compact schedule spaces, validates generated schedules through LLVM execution and HLS C/C++ code generation, and ranks valid candidates before Vitis HLS synthesis. Rather than replacing vendor HLS tools, analytical models, or learning-based DSE methods, \tool{} acts as a front-end pruning layer that forwards fewer and more relevant candidates to expensive synthesis.

The contributions of this paper are as follows:
\begin{itemize}
\item We formulate pattern-guided schedule-space pruning as a practical HLS DSE layer for recurring numerical kernels.
\item We implement an end-to-end workflow that connects pattern-aware candidate generation, backend validation, estimator-based ranking, and Vitis HLS evaluation through the \allo{} scheduling interface.
\item We evaluate six kernels and show that \tool{} reduces HLS-evaluated candidates from 140 to 29 while recovering the same best valid Vitis HLS latency found by the exhaustive-lite baseline.
\end{itemize}

\section{Background and Motivation}
\label{sec:background}

\subsection{Schedule Spaces in HLS}
An HLS design is determined not only by the functional kernel but also by a set of schedule decisions. A candidate design can be represented as
\begin{equation}
    d = (p, u, r, t, m, b, \ldots),
\end{equation}
where $p$ denotes pipeline choices, $u$ denotes unroll factors, $r$ denotes loop reordering, $t$ denotes tiling, $m$ denotes memory partitioning, and $b$ denotes buffering decisions. These knobs interact with each other. Increasing an unroll factor can expose parallelism, but it may also increase memory-port pressure or resource usage. Pipelining a loop can reduce initiation interval, but loop-carried dependencies may prevent the intended schedule. Tiling can improve locality, but it may add local storage and control overhead.

Compiler frameworks such as ScaleHLS and scheduling languages such as Halide, TVM, HeteroCL, and \allo{} demonstrate the value of separating what a program computes from how it is scheduled \cite{scalehls,halide,tvm,heterocl,allo}. However, this separation also shifts part of the optimization burden to DSE. Even when each individual transformation is well-defined, their composition creates a large search space.

\subsection{Allo-Style HLS Schedule Construction}
Figure~\ref{lst:allo_gemm} shows a simplified Allo-style GEMM fragment adapted from the schedule-construction style in \cite{allo}. The algorithm specification defines the loop nest and computation, while the schedule object applies transformations such as pipelining, unrolling, and array partitioning. In this setting, \tool{} does not change the programming model. Instead, it decides which schedule primitives and parameter values should be instantiated for a given computation pattern.

\begin{figure}[t]
\begin{lstlisting}[style=allopy]
def gemm(A: int8[M, K], B: int8[K, N],
         C: int16[M, N]):
    for i, j in allo.grid(M, N, name="C"):
        for k in range(K):
            C[i, j] += A[i, k] * B[k, j]

s = allo.customize(gemm)
s.pipeline("k")
s.unroll("j", factor=UF)
s.partition(s.A, dim=1, factor=UF)
s.partition(s.B, dim=0, factor=UF)
s.partition(s.C, dim=1, factor=UF)
\end{lstlisting}
\caption{Simplified Allo-style GEMM kernel and schedule fragment. The schedule choices expose the kind of search knobs that \tool{} constrains using computation-pattern information.}
\label{lst:allo_gemm}
\end{figure}

\subsection{Computation Patterns as DSE Structure}
Many FPGA kernels exhibit recognizable computational structure. Elementwise kernels have independent iterations and limited data reuse. Reduction kernels contain accumulation dependencies and benefit from reduction-safe unrolling or tree-like accumulation. Matrix-vector and matrix-matrix kernels have structured reuse across rows, columns, and reduction axes. Stencil-like kernels access neighboring elements and often benefit from locality-aware scheduling and buffering.

These differences imply that a single generic search template is unlikely to be efficient across all kernels. For a simple elementwise map, exploring many loop permutations or tiling choices is unnecessary. For matrix multiplication, however, loop order and unroll-axis selection strongly affect reuse and memory bandwidth. For stencil updates, spatial locality matters more than blindly unrolling every loop. Pattern information therefore provides a natural prior over useful schedules.

% \subsection{Design Objective}
% The objective of \tool{} is to reduce the number of candidates sent to HLS synthesis while preserving high-quality outcomes. This objective matters because downstream FPGA evaluation is not only a compiler problem: candidate quality is ultimately affected by HLS scheduling, memory banking, placement, routing, and timing closure. The framework is intentionally lightweight: it avoids full polyhedral analysis and does not require training data. Instead, it uses structural metadata to classify kernels and choose a compact set of candidate schedules. This makes the DSE process easier to debug and easier to extend with new pattern templates.

\section{System Overview}
\label{sec:overview}

\begin{figure}[t]
\centering
\includegraphics[width=\linewidth]{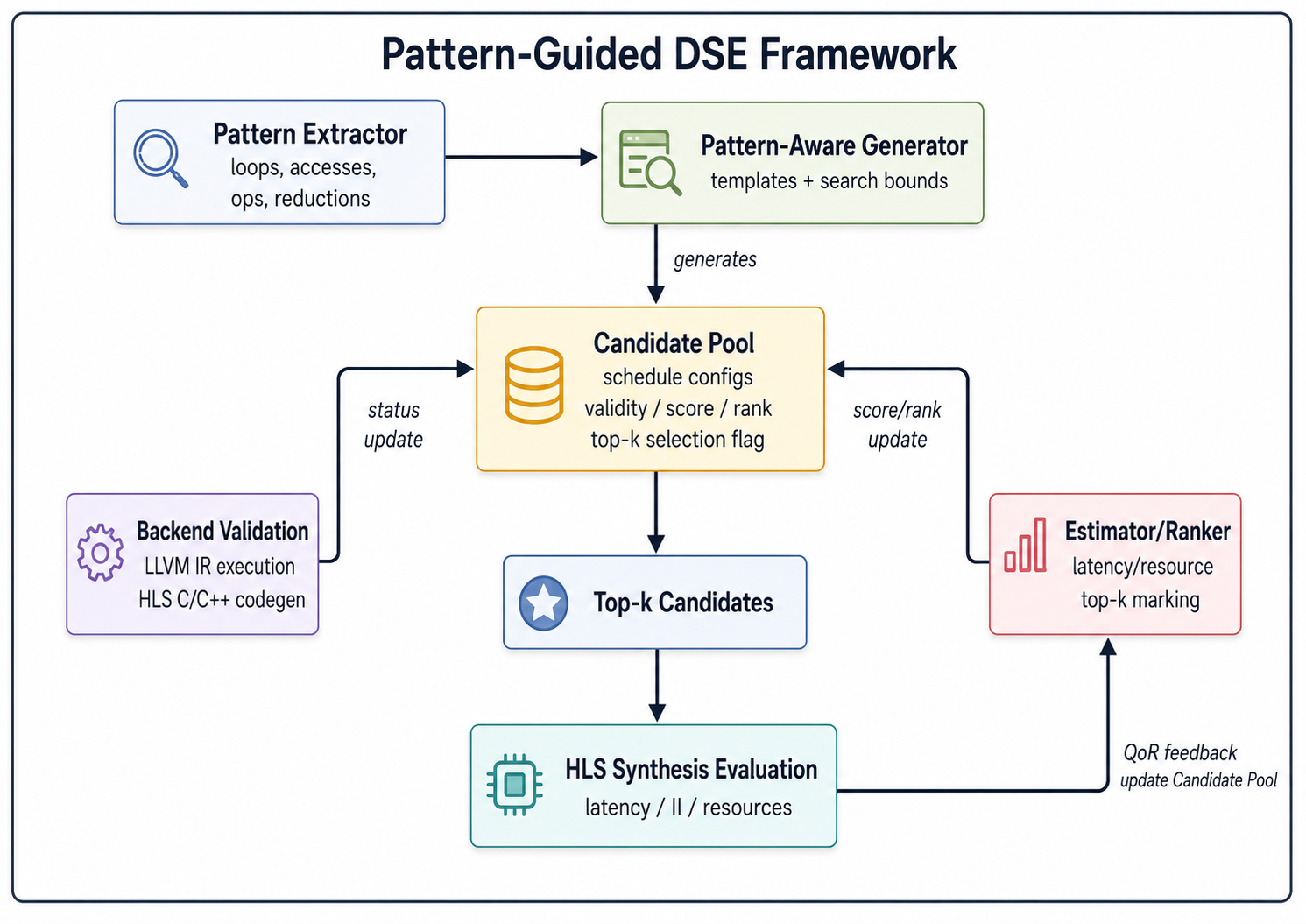}
\caption{Overview of \tool{} workflow.}
\label{fig:overview}
\end{figure}

Fig.~\ref{fig:overview} shows the \tool{} workflow. The input is a kernel written in \allo{} and a bounded search configuration. In the current prototype, each benchmark kernel is associated with a computation pattern and a small amount of schedule metadata, such as candidate loop axes, reduction axes, and search bounds.

The pattern-aware generator maps this pattern information to schedule templates and search bounds. Rather than enumerating every combination of generic schedule knobs, it generates candidates that match the selected pattern template. The candidate pool stores schedule configurations together with validity, score, rank, and top-$k$ selection status. Backend validation applies each schedule, checks LLVM-level execution, and verifies that HLS C/C++ code can be generated. The estimator/ranker then updates the candidate pool using latency and resource estimates. Finally, the selected top-$k$ candidates are evaluated through Vitis HLS synthesis to obtain latency, initiation interval (II), and resource usage.

The feedback loop in Fig.~\ref{fig:overview} is important. In the current prototype, the estimator is a simple pattern-aware heuristic used to prioritize candidates before synthesis. The same interface can later support calibrated analytical models, synthesis-history feedback, or learning-based rankers without changing the rest of the flow.

\section{Pattern-Guided DSE Method}
\label{sec:method}

\subsection{Pattern Templates}
In this prototype, \tool{} supports five pattern classes: \texttt{elementwise}, \texttt{reduction}, \texttt{matvec}, \texttt{gemm}, and \texttt{stencil}. The current implementation does not rely on a complicated standalone extractor. Instead, each supported kernel is mapped to a pattern template that specifies the schedule primitives and search bounds worth considering first. Kernels that do not match the supported templates can fall back to a generic bounded search.

Table~\ref{tab:patterns} summarizes the pattern templates used in the current implementation. The templates are deliberately compact: they encode useful schedule priors without trying to prove every legal transformation before backend validation.

\begin{table}[t]
\caption{Pattern Templates Used by \tool{}}
\label{tab:patterns}
\centering
\footnotesize
\begin{tabular}{p{0.20\linewidth}p{0.25\linewidth}p{0.44\linewidth}}
\toprule
Pattern & Kernels & Schedule focus \\
\midrule
Elementwise & \texttt{vecadd}, \texttt{axpy} & Pipeline the main loop and try small unroll factors. Avoid tiling and loop permutations that do not expose reuse. \\
Reduction & \texttt{dot} & Pipeline the reduction loop and test reduction-safe unroll factors while preserving accumulation correctness. \\
MatVec & \texttt{matvec} & Select between output-axis and reduction-axis parallelism under memory-bandwidth constraints. \\
GEMM & \texttt{gemm} & Explore selected loop orders and bounded unroll combinations that expose multiply-accumulate parallelism. \\
Stencil & \texttt{jacobi2d} & Pipeline spatial loops and keep the search compact for locality-oriented extensions. \\
\bottomrule
\end{tabular}
\end{table}

% The previous standalone pattern-extraction algorithm is intentionally omitted.
% The current prototype focuses on compact pattern templates and backend validation.

\subsection{Candidate Generation, Validation, and Ranking}
Algorithm~\ref{alg:dse} describes the full DSE loop. The pattern class and optional schedule metadata select a rule set from the pattern library. The generator instantiates only bounded schedules allowed by that rule set. Each schedule is applied through the \allo{} schedule interface. Candidates that fail LLVM execution or HLS C/C++ code generation are removed before synthesis. Valid candidates are scored as
\begin{equation}
S(d) = \widehat{L}(d) \cdot \left(1 + \alpha \widehat{R}(d)\right),
\label{eq:score}
\end{equation}
where $\widehat{L}(d)$ estimates latency, $\widehat{R}(d)$ estimates normalized resource pressure, and $\alpha$ controls the resource penalty. This score is used only to prioritize candidates. Final quality is still measured by Vitis HLS.

\begin{algorithm}[t]
\caption{Pattern-Guided Candidate Generation and Ranking}
\label{alg:dse}
\footnotesize
\begin{algorithmic}[1]
\REQUIRE Kernel $K$, pattern class $c$, optional metadata $M$, rule library $\mathcal{R}$, budget $B$
\ENSURE Ranked candidate list $C_{ranked}$
\STATE $C \leftarrow \emptyset$, $C_{valid} \leftarrow \emptyset$
\STATE Select pattern rules $R_c \leftarrow \mathcal{R}[c]$
\FOR{each template $r \in R_c$}
    \STATE Instantiate schedule parameters within budget $B$
    \STATE Append generated schedules to $C$
\ENDFOR
\FOR{each candidate $d \in C$}
    \STATE Apply $d$ to the schedule object
    \IF{$d$ passes LLVM execution}
        \IF{$d$ emits HLS C/C++ code}
            \STATE Estimate $\widehat{L}(d)$ and $\widehat{R}(d)$
            \STATE Add $(d,S(d))$ to $C_{valid}$ using Eq.~\eqref{eq:score}
        \ENDIF
    \ENDIF
\ENDFOR
\STATE Sort $C_{valid}$ by score and mark the top-$k$ candidates
\RETURN $C_{ranked}$
\end{algorithmic}
\end{algorithm}

This formulation gives \tool{} three practical advantages. First, it reduces the initial schedule space before expensive HLS synthesis. Second, it keeps generated candidates interpretable because each candidate is associated with a pattern rule. Third, it provides an extension path: supporting a new kernel family only requires adding a pattern rule or mapping it to an existing pattern.

\section{Implementation}
\label{sec:impl}

We implemented \tool{} as a Python-based DSE layer around the \allo{} workflow. The prototype contains benchmark kernels, pattern metadata, schedule templates, validation scripts, ranking utilities, and Vitis HLS batch-evaluation scripts. The implementation separates four concerns: search-space construction, schedule application, backend validation, and candidate ranking.

The evaluation pipeline proceeds as follows. First, a counting script enumerates the exhaustive-lite and pattern-guided spaces under the same bounds. Second, a DSE script applies each generated schedule and records whether it passes LLVM execution and HLS C/C++ code generation. Third, a ranking script computes estimator scores and marks the selected candidates. Fourth, a batch HLS script synthesizes exported candidates with the same Vitis HLS configuration.

All HLS evaluations use the same target part and clock setting within an experiment. The exported projects are organized by kernel, method, and candidate identifier so that exhaustive-lite and pattern-guided candidates can be compared using the same downstream synthesis flow. This organization also makes it possible to rerun only selected candidates when the ranking heuristic changes.

\section{Evaluation}
\label{sec:eval}

\subsection{Experimental Setup}
We evaluate \tool{} on six kernels: \texttt{vecadd}, \texttt{axpy}, \texttt{dot}, \texttt{matvec}, \texttt{gemm}, and \texttt{jacobi2d}. These kernels cover elementwise computation, scalar-vector operations, reductions, matrix-vector computation, matrix multiplication, and stencil-like updates. The baseline is \emph{exhaustive-lite}, a bounded enumeration strategy that applies a generic schedule template to each kernel. Exhaustive-lite is not meant to represent a full unbounded exhaustive search; it is a controlled baseline that exposes how many candidates are produced when pattern information is not used.

We report three metrics. First, candidate count measures how many designs are sent to HLS evaluation. Second, search-space reduction measures the percentage reduction relative to exhaustive-lite. Third, best valid HLS latency measures whether the smaller pattern-guided space preserves the best synthesized latency found by the baseline. A candidate is considered valid only if it passes backend checks and produces a Vitis HLS result.

\subsection{Search-Space Reduction}
Table~\ref{tab:space} shows the candidate counts. Across the six kernels, exhaustive-lite evaluates 140 candidates, while \tool{} evaluates 29 candidates. This corresponds to a 4.83$\times$ overall reduction. The reduction is largest for kernels with richer nested-loop structure: \texttt{gemm} is reduced from 72 to 12 candidates, and \texttt{jacobi2d} is reduced from 36 to 3 candidates.

% \begin{table}[t]
% \caption{Candidate Count Reduction}
% \label{tab:space}
% \centering
% \footnotesize
% \begin{tabular}{lrrrr}
% \toprule
% Kernel & Pattern & Exhaustive-lite & \tool{} & Reduction \\
% \midrule
% \texttt{vecadd} & Elementwise & 8 & 4 & 2.00$\times$ \\
% \texttt{axpy} & Elementwise & 8 & 4 & 2.00$\times$ \\
% \texttt{dot} & Reduction & 8 & 3 & 2.67$\times$ \\
% \texttt{matvec} & MatVec & 8 & 3 & 2.67$\times$ \\
% \texttt{gemm} & GEMM & 72 & 12 & 6.00$\times$ \\
% \texttt{jacobi2d} & Stencil & 36 & 3 & 12.00$\times$ \\
% \midrule
% Total & -- & 140 & 29 & 4.83$\times$ \\
% \bottomrule
% \end{tabular}
% \end{table}

\begin{table}[t]
\caption{Candidate Count Reduction}
\label{tab:space}
\centering
\footnotesize
\setlength{\tabcolsep}{3pt}
\begin{tabular}{@{}llrrr@{}}
\toprule
Kernel & Pattern & Exhaustive-lite & \tool{} & Reduction \\
\midrule
\texttt{vecadd} & Elementwise & 8 & 4 & 2.00$\times$ \\
\texttt{axpy} & Elementwise & 8 & 4 & 2.00$\times$ \\
\texttt{dot} & Reduction & 8 & 3 & 2.67$\times$ \\
\texttt{matvec} & MatVec & 8 & 3 & 2.67$\times$ \\
\texttt{gemm} & GEMM & 72 & 12 & 6.00$\times$ \\
\texttt{jacobi2d} & Stencil & 36 & 3 & 12.00$\times$ \\
\midrule
Total & -- & 140 & 29 & 4.83$\times$ \\
\bottomrule
\end{tabular}
\end{table}

\begin{figure*}[t]
\centering
\begin{minipage}{0.48\textwidth}
\centering
\includegraphics[width=\linewidth]{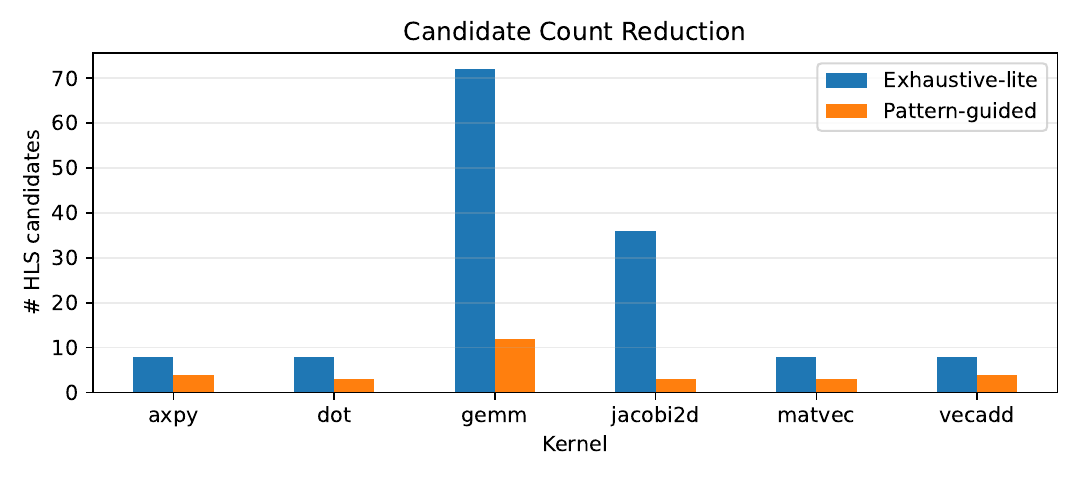}
\end{minipage}
\hfill
\begin{minipage}{0.48\textwidth}
\centering
\includegraphics[width=\linewidth]{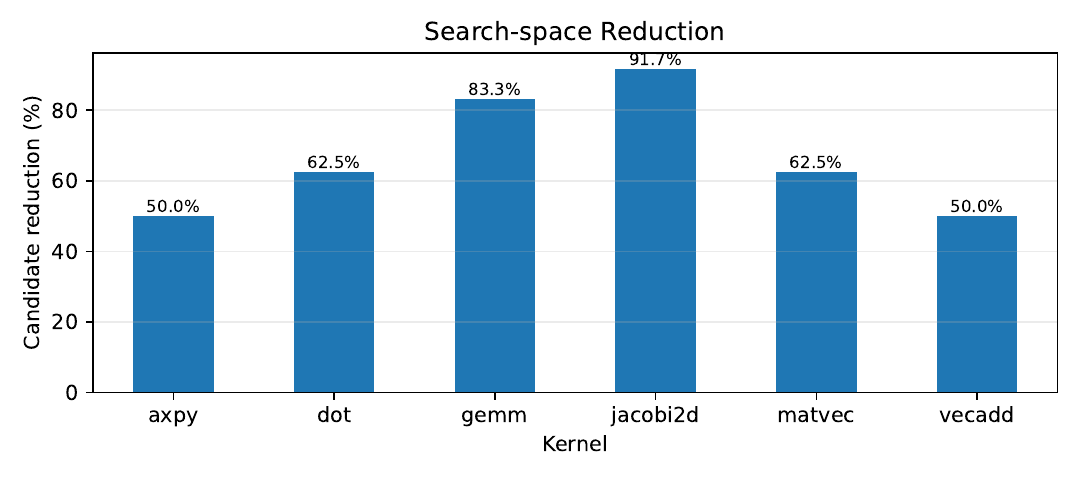}
\end{minipage}
\caption{Search-space reduction results. Left: number of HLS candidates evaluated by exhaustive-lite and \tool{}. Right: percentage reduction achieved by pattern-guided candidate generation.}
\label{fig:count_reduction}
\end{figure*}

Fig.~\ref{fig:count_reduction} visualizes both the absolute candidate counts and the corresponding percentage reduction. The smaller reductions for \texttt{vecadd} and \texttt{axpy} are expected because their baseline spaces are already small. The larger reductions for \texttt{gemm} and \texttt{jacobi2d} show the advantage of using computation structure when generic schedule knobs become combinatorial.

\subsection{Best Valid HLS Latency}
\begin{figure}[t]
\centering
\includegraphics[width=\linewidth]{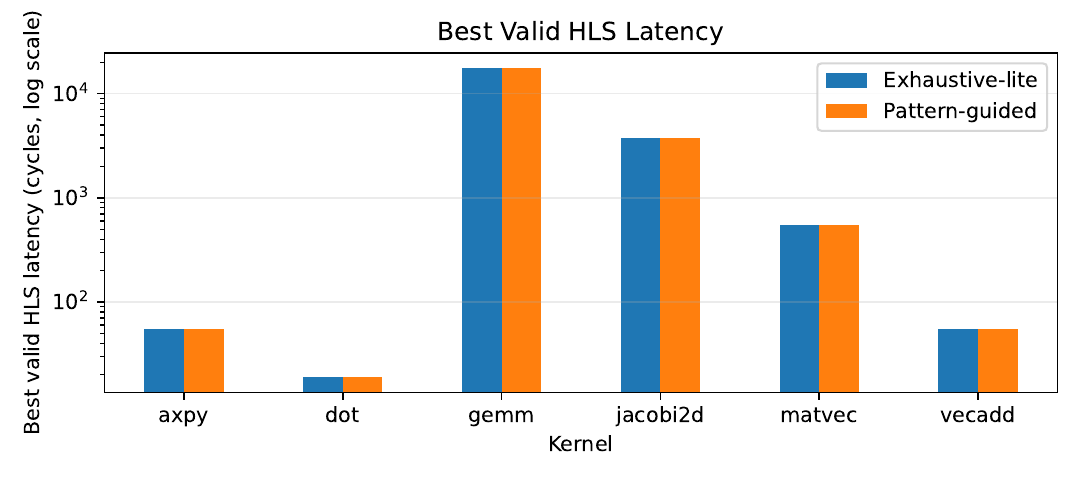}
\caption{Best valid Vitis HLS latency found by exhaustive-lite and \tool{}. Pattern-guided search recovers the same best valid latency while evaluating fewer candidates. The y-axis uses a logarithmic scale.}
\label{fig:best_latency}
\end{figure}

The key question is whether pruning removes useful candidates. Fig.~\ref{fig:best_latency} compares the best valid Vitis HLS latency found by exhaustive-lite and \tool{} for each kernel. Across all six kernels, the best valid latency found by \tool{} matches the best valid latency found by exhaustive-lite. This result indicates that the removed candidates are not necessary for reaching the best latency within the evaluated search bounds.

\subsection{Discussion}
The results support the central hypothesis that computation patterns can guide HLS schedule exploration. Pattern guidance is most beneficial when the generic schedule space contains many combinations that are structurally legal but unlikely to be productive. For example, matrix multiplication and stencil-like updates have more loop axes and more possible schedule choices than simple elementwise kernels; as a result, \tool{} removes a larger fraction of candidates for these kernels.

The current prototype emphasizes pruning and ranking rather than aggressive schedule invention. This choice is deliberate. By leaving final quality evaluation to Vitis HLS, \tool{} remains compatible with standard HLS reporting and avoids overfitting to an internal cost model. The estimator only affects which candidates are prioritized. In future work, synthesis feedback can be used to calibrate the estimator and improve top-$k$ selection.

\section{Related Work}
\label{sec:related}

\subsection{FPGA Accelerator Workloads and AI Systems}
FPGA accelerator research is increasingly shaped by workload and deployment context. Cloud systems such as Catapult and later datacenter FPGA deployments demonstrate how reconfigurable logic can be integrated into production infrastructure, where latency, throughput, resource cost, and flexibility must be considered together \cite{putnam_catapult,caulfield_cloud_scale}. Benchmark suites such as MachSuite, Rodinia, CHStone, and Rosetta provide representative workloads for customized architectures, heterogeneous computing, and HLS research \cite{machsuite,rodinia,chstone,rosetta}. AI further highlights the need for workload-aware design: MLPerf standardizes inference evaluation \cite{mlperf_inference}, FPGA-oriented frameworks such as fpgaConvNet and DNNBuilder automate neural-network accelerator construction \cite{fpgaconvnet,dnnbuilder}, and systems such as TPU and Gemmini show the value of matching architecture, software, and workload structure across the full stack \cite{tpu,gemmini}. \tool{} targets a narrower layer in this landscape: it uses workload pattern information to constrain HLS schedule-space exploration.

\subsection{HLS Programming Systems and FPGA Implementation Flows}
HLS has evolved from C/C++-oriented FPGA compilation systems such as LegUp into a broader ecosystem of scheduling-oriented and domain-specific programming flows \cite{legup,hls_survey}. Halide, TVM, HeteroCL, ScaleHLS, and \allo{} all separate the algorithm from scheduling or mapping decisions in different domains \cite{halide,tvm,heterocl,scalehls,allo}, while MLIR provides reusable compiler infrastructure for building such domain-specific flows \cite{mlir}. At the same time, FPGA implementation quality depends on backend effects such as memory banking, placement, routing, and timing closure. VTR provides open FPGA CAD infrastructure \cite{vtr}, and AutoBridge and TAPA show that high-frequency FPGA designs often require coupling HLS-level transformations with floorplanning, pipelining, dataflow organization, and physical-design awareness \cite{autobridge,tapa}. Hardware construction languages and open implementation flows, including Chisel and OpenROAD, reflect the same broader trend toward generator-based design and automation \cite{chisel,openroad}. \tool{} is complementary to these systems: it filters schedule candidates before expensive backend evaluation.

\subsection{DSE, Estimation, and Pattern-Guided Search}
HLS DSE has been studied through analytical models, heuristic search, learning-based methods, and pragma optimization frameworks \cite{schafer_hls_dse,liu_learning_hls,autodse, 202607.0874}. These methods seek to reduce synthesis cost while still finding high-quality implementations. \tool{} does not propose a new global optimizer; instead, it reduces the initial candidate pool using computation-pattern information, then relies on backend validation and Vitis HLS to measure final quality. This design is motivated by the fact that programs exhibit recurring structures, as captured by the Berkeley dwarfs \cite{asanovic_dwarfs}. Reductions, dense linear algebra, stencils, convolutions, and neural-network layers each have distinct dependency and data-reuse behavior, so mapping them to the same generic schedule template can waste evaluation budget. \tool{} applies this idea at the schedule-space level by mapping recurring kernel patterns to compact schedule templates and using a simple estimator/ranker to prioritize valid candidates.

\section{Limitations and Future Work}
\label{sec:future}

The current prototype has several limitations. First, the pattern library is manually defined and covers a small set of numerical kernels. Future work should add convolutions, tensor contractions, sparse kernels, and streaming dataflow patterns. Second, the estimator is intentionally simple. More accurate analytical models or learning-based rankers could improve top-$k$ selection, especially when the pattern-guided candidate set becomes larger. Third, the current templates focus on loop scheduling and bounded unrolling; more advanced memory hierarchy generation, tiling, and buffering strategies should be incorporated. Finally, the evaluation currently compares against an exhaustive-lite baseline. Stronger baselines, including Bayesian optimization, genetic search, and prior HLS DSE tools, would further clarify where pattern guidance provides the largest benefit.

Despite these limitations, the results suggest that pattern-guided pruning is a useful design point. It preserves the transparency of schedule-based programming while avoiding unnecessary downstream synthesis of structurally mismatched candidates.

\section{Conclusion}
\label{sec:conclusion}

This paper presented \tool{}, a lightweight pattern-guided DSE framework for FPGA kernels written in \allo{}. \tool{} uses computation structure to instantiate compact schedule spaces, validate candidate schedules, rank valid candidates, and evaluate selected designs through Vitis HLS. On six representative kernels, \tool{} reduces the number of HLS-evaluated candidates from 140 to 29, achieving a 4.83$\times$ overall search reduction and up to 12.0$\times$ reduction for an individual kernel. At the same time, it recovers the same best valid Vitis HLS latency as the exhaustive-lite baseline for all evaluated kernels. These results show that simple computation-pattern information can make HLS schedule exploration more efficient without sacrificing the best observed HLS outcome within the evaluated search bounds. Realizing this at scale will require shared benchmarks that evaluate DSE methods through executable synthesis rather than self-defined search bounds \cite{NEURIPS2025_664f7775}.

\bibliographystyle{IEEEtran}
\bibliography{refs}

\end{document}